\documentclass[conference]{IEEEtran}
\usepackage{graphicx} 
\usepackage{hyperref}
\usepackage{comment}
\usepackage[disable]{todonotes}
\usepackage{enumitem}
\usepackage[icelandic,english]{babel}

\hypersetup{
    colorlinks=true,
    linkcolor=black,
    filecolor=black,      
    citecolor=black,
    urlcolor=blue
}
            
\title{Automated Configuration Synthesis for Machine Learning Models: A git-Based Requirement and Architecture Management System}
\author{
\IEEEauthorblockN{Abdullatif AlShriaf}
\IEEEauthorblockA{Chalmers and University of Gothenburg\\
Gothenburg, Sweden\\
Email: shriaf@chalmers.se}
\and
\IEEEauthorblockN{Hans-Martin Heyn}
\IEEEauthorblockA{Chalmers and University of Gothenburg\\
Gothenburg, Sweden\\
Email: heyn@chalmers.se}
\and
\IEEEauthorblockN{Eric Knauss}
\IEEEauthorblockA{Chalmers and University of Gothenburg\\
Gothenburg, Sweden\\
Email: eric.knauss@cse.gu.se}
}
\date{March 2024}

\usepackage{ifthen}
\usepackage{amssymb}
\newboolean{showcomments}
\setboolean{showcomments}{true} 

\ifthenelse{\boolean{showcomments}}
  {\newcommand{\nb}[2]{
    \fcolorbox{gray}{yellow}{\bfseries\sffamily\scriptsize#1}
    {\sf\small$\blacktriangleright$\textit{#2}$\blacktriangleleft$}
   }
   
  }
  {\newcommand{\nb}[2]{}
   
  }

\usepackage[normalem]{ulem} 
\usepackage{xcolor}

\ifthenelse{\boolean{showcomments}}
  {
  
  \newcommand{\ins}[1]{\textcolor{blue}{\uline{#1}}} 
  \newcommand{\del}[1]{\textcolor{red}{\sout{#1}}} 
  \newcommand{\inspar}[1]{\color{blue}{#1}\color{blue}} 
  }
  {
  
  \newcommand{\ins}[1]{#1} 
  \newcommand{\del}[1]{} 
  \newcommand{\inspar}[1]{} 
  
  }

\begin{document}

\maketitle

\section{Background}
\todo[inline]{Things that I feel would be really interesting to write about in this context:
\newline
- The idea of extracting data from semi-formal requirements. This will make it more attractive to keep requirements updated with code.
\newline
- The idea to keep requirements and corresponding architectural decisions aligned with code in git.
\newline
But I am not so really sure what is the key message that we have? (Eric)
}

\todo[inline]{We suggest the following format for the extended abstract (2 pages max).\newline
Context (including, e.g., users or stakeholders of the proposed tool/method/technique); \newline
RE problem (challenge) addressed and Motivation; \newline
Methodology used or the tool you developed; \newline
any Validation/Evaluation conducted or planned; \newline
Results and Contributions, \newline
and a discussion of Related Work \newline
\newline
Optional and recommended: \newline
A set of up to 3 questions that the authors would like the attendees of the conference to answer to help them validate their work. The questions shall be complemented with a short description on why these questions are important and how the answers would help in the further work. During the Tool and Demo session, each stand will have a box and printed-out questions to answer them on a paper, as well as links to an electronic form to answer the questions electronically.
}
\todo[inline]{An annex to the abstract (max 2 pages) with: \newline
For posters and student research posters : \newline
A draft of a poster to which a short description explaining clearly how the poster is to be presented at the conference can be added.\newline
For tool demos: \newline
A short description or a link to video explaining clearly how the work described in the extended abstract is to be presented at the conference. The annex should emphasize interaction potential. It may contain images of screenshots or URLs leading to more information about the work, e.g., a video clip.}

The design of complex distributed systems typically follows a hierarchical process, supported by highly specialized views for decomposing the design task. Requirements and architecture often evolve simultaneously, requiring an architectural framework that supports integrated and collaborative design, including non-functional requirements and quality views. The framework must ensure the traceability of design decisions in order to build safety cases.
Integrating requirements into software development is vital for aligning intended functionality with implemented code.
However, extracting data from semi-formal requirements and maintaining alignment poses challenges due to its ambiguity and variability making extracting consistent information challenging. Aligning these requirements with other project artifacts can also be difficult due to interpretation differences, often requiring manual effort and leading to complexity and potential inconsistencies in development \cite{SEMI_FORMAL}.
\section{Introduction}
\begin{figure}
    \centering
    \includegraphics[width=0.5\textwidth]{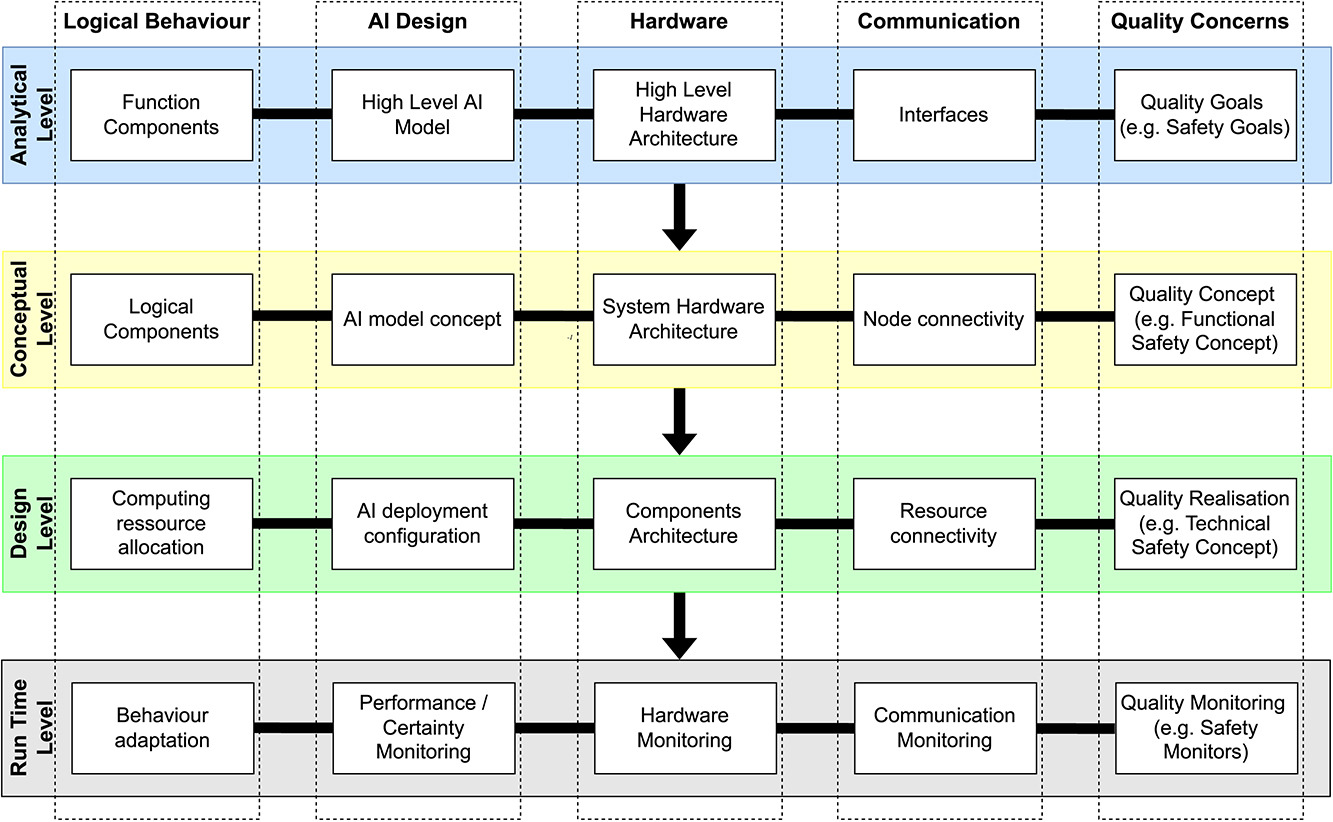}
    \caption{The VEDLIoT Architectural Framework \cite{Heyn2023}.}
    \label{fig:vedliot-af}
\end{figure}

We propose GRAMS\footnote{https://gitlab.com/latiif/configgen} as a git-based Requirements and Architecture Management System which we demonstrate can automate configuration extraction, reduce manual effort, enhance configurability, and simplify runtime configuration management. 
We will show that it improves traceability by establishing clear links between requirements, configurations, and code, and facilitates maintenance through a structured approach. 
GRAMS is intended for use by i) machine learning engineers and architects tasked with designing the structure of ML-based software systems, and ii) system administrators/MLOps engineers responsible for deploying and managing the runtime environment of software systems.
GRAMS builds upon the open-source tool \emph{T-Reqs} for managing textual requirements in git \cite{treqs_knauss}. 
It extends T-Reqs by implementing a structured architectural view system based on the concept of a compositional architectural framework for the development of AI-enabled software systems (Figure \ref{fig:vedliot-af},  \cite{Heyn2023}). Using a schema-type template, GRAMS compiles a machine-readable YAML \emph{intermediary} file to communicate AI model specifications. This file can then be forwarded and converted as part of a deployment-toolchain, particularly for example with the purpose of AI model optimization.
\todo[inline]{If we present both GRAMS and T-Reqs, we need to be very clear about the differences. 
Is GRAMS simply the implementation of the compositional architectural framework in T-Reqs? Since we are then not arguing with agile, what other argument do we have that git is a good idea? (Eric)\newline
(latiif): The tool GRAMS operates on T-Reqs-implementations of Arch. Framework: Identifies explicit runtime-scenarios, Heurestically traverses the AF, Locates configurations/schemes \& maps them to the runtime-scenarios.\newline
For me, GRAMS is the extended TTIM (2 extra types) + Traversal Algorithm + Aggregation(as YAML or PlantUML) } 

\section{Context: Compositional Architectural Framework}
\label{sec:framework}
The system design approach supported by GRAMS follows a scalable and compositional architectural framework that provides flexibility to manage requirements in software projects of varying complexity. 
The architectural framework organizes quality concerns, including ethical considerations such as explainability and security. 
Supporting middle-out design, GRAMS integrates existing design decisions, enhancing alignment between requirements and architecture \cite{Heyn2023}. Guiding rules ensure traceability, vital for managing requirements throughout development. The framework's adaptability to diverse abstraction levels enables GRAMS to cater to varied project needs.
\section{T-Reqs: Text-based Requirements Management System}
GRAMS extends T-Reqs, an open-source tool for the management of textual requirements in git.
T-Reqs is developed at Chalmers and University of Gothenburg and used in industrial settings, including at Ericsson \cite{EVAL_TEXTBASED}. 
T-Reqs utilizes a TTIM, or Type and Trace Information Model, that extends the functionality of the basic Traceability Information Model (TIM) by incorporating type information. While a TIM primarily focuses on representing trace-links and relationships between abstract trace artifacts~\cite{TRA}, the TTIM defines types of traced elements and their connections, which can be optional or required~\cite{treqs_knauss}. T-Reqs operates as a static analysis tool on the model described by the TTIM.
\section{Applying GRAMS}
\begin{figure}
    \centering
    \includegraphics[width=0.5\textwidth]{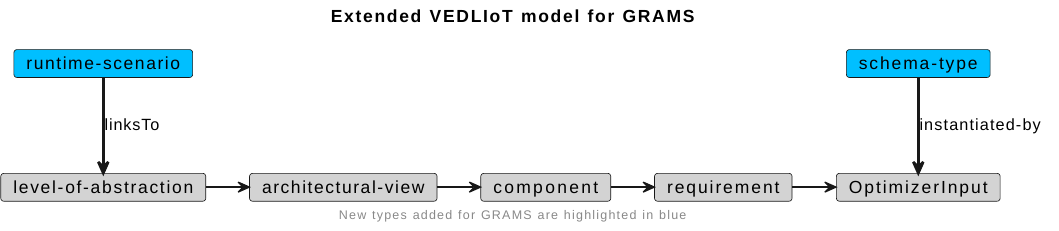}
    \caption{Extended Architectural Framework for GRAMS: New types added for GRAMS are highlighted in blue} 
    \label{fig:addition_to_caf}
\end{figure}

\begin{figure}
    \centering
    \includegraphics[width=0.5\textwidth]{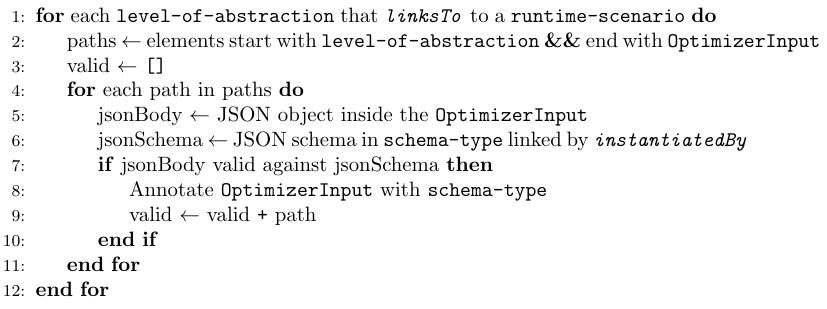}
    \caption{Framework traversal algorithm for GRAMS in pseudo-code.}
    \label{fig:grams_workflow}
\end{figure}

First, GRAMS identifies explicit runtime scenarios, then heuristically traverses the architectural framework, and finally locating configurations/schemes and mapping them to the identified runtime scenarios.
In essence, GRAMS executes the TTIM, as illustrated in Figure~\ref{fig:addition_to_caf}, along with a traversal algorithm which is outlined in Figure~\ref{fig:grams_workflow}. Finally, the aggregation capabilities of the system are presented in generic formats such as YAML or PlantUML.
Leveraging the architectural structure laid by the framework, we introduce two new elements, as shown in Figure \ref{fig:addition_to_caf}:
\begin{itemize}[leftmargin=3mm] 
    \item \textit{runtime-scenario}: matches particular models to specific contexts and links to a level of abstraction which GRAMS will scan and traverse all the way to an AI-model optimiser input referred to as \textit{OptimizerInput}.
    \item \textit{schema-type}: Annotates the \textit{OptimizerInput} and contains the actual schema for the JSON object present in the instantiating \textit{OptimizerInput}.
\end{itemize}

Multiple \textit{runtime-scenario}s can exist in the framework. From a single \textit{runtime-scenario}, multiple \textit{OptimizerInputs} can be traced.
The traversal follows the algorithm outlined in Figure \ref{fig:grams_workflow}.

\section{Quality Assurance Checks in GRAMS}
\label{sec:checks}
To ensure output quality and usability, GRAMS conducts the following checks:
\begin{itemize}[leftmargin=3mm] 
    \item Consistency is crucial to prevent conflicts and maintain coherence \cite{lauesen_2002}. The initial "dumb" check ensures alignment with the framework's meta-model (see Section \ref{sec:framework}).
    \item GRAMS validates the internal correctness of the configuration-relevant requirements specification by checking JSON schema compliance with \textit{schema-type} and \textit{OptimizerInput}.
    \item GRAMS verifies semantic equivalence between relevant parts of the configuration schema and \textit{OptimizerInput} schemas to confirm configuration relevance to \textit{runtime-scenarios}.
\end{itemize}

\section{Discussion \& Conclusion}
Having demonstrated the feasibility of leveraging the architectural framework for specific runtime configurations, the necessity for explicit information about dynamic property schema becomes apparent. Integrating architecture with requirement context facilitates traceable configuration generation, anchored in system design and requirements. GRAMS' scalability and reliance on the architectural framework provide a smoother topology for traversing all possible variable inputs.
By specifying the requirements for the system and each model in every runtime scenario, we enable traceability and support safety argumentation. Additionally, generating configurations for the model optimizer from these requirements facilitates frequent changes and iterative development, necessitating the continuous upkeep of requirements to ensure alignment between the optimized models and their specifications.
Yet, questions arise: Is the intermediary format sufficient for tracing relevant properties and generating final configurations? Are GRAMS checks overly strict, and what adjustments are needed? Additionally, should GRAMS be extended to generic systems not reliant on the compositional architectural framework? These queries prompt further examination. 

\bibliographystyle{IEEEtran}
\bibliography{references}

\begin{thebibliography}{1}
\providecommand{\url}[1]{#1}
\csname url@samestyle\endcsname
\providecommand{\newblock}{\relax}
\providecommand{\bibinfo}[2]{#2}
\providecommand{\BIBentrySTDinterwordspacing}{\spaceskip=0pt\relax}
\providecommand{\BIBentryALTinterwordstretchfactor}{4}
\providecommand{\BIBentryALTinterwordspacing}{\spaceskip=\fontdimen2\font plus
\BIBentryALTinterwordstretchfactor\fontdimen3\font minus
  \fontdimen4\font\relax}
\providecommand{\BIBforeignlanguage}[2]{{%
\expandafter\ifx\csname l@#1\endcsname\relax
\typeout{** WARNING: IEEEtran.bst: No hyphenation pattern has been}%
\typeout{** loaded for the language `#1'. Using the pattern for}%
\typeout{** the default language instead.}%
\else
\language=\csname l@#1\endcsname
\fi
#2}}
\providecommand{\BIBdecl}{\relax}
\BIBdecl

\bibitem{SEMI_FORMAL}
A.~Zaki-Ismail, M.~Osama, M.~Abdelrazek, J.~Grundy, and A.~Ibrahim,
  ``\BIBforeignlanguage{English}{Requirements formality levels analysis and
  transformation of formal notations into semi-formal and informal
  notations},'' ser. Proc. of the Int. Conf. on Software Engineering and
  Knowledge Engineering, SEKE.\hskip 1em plus 0.5em minus 0.4em\relax KSI
  Research Inc., 2021, pp. 303--308.

\bibitem{treqs_knauss}
E.~Knauss, G.~Liebel, J.~Horkoff, R.~Wohlrab, R.~Kasauli, F.~Lange, and
  P.~Gildert, ``T-reqs: Tool support for managing requirements in large-scale
  agile system development,'' in \emph{2018 IEEE 26th International
  Requirements Engineering Conference (RE)}, 2018, pp. 502--503.

\bibitem{Heyn2023}
H.-M. Heyn, E.~Knauss, and P.~Pelliccione, ``A compositional approach to
  creating architecture frameworks with an application to distributed ai
  systems,'' \emph{Systems and Software (JSS)}, vol. 198, 2023.

\bibitem{EVAL_TEXTBASED}
N.~Theod\'orsd\'ottir, Audur Katar\'ina~Ojensa, \emph{Evaluation of
  Text–Based Requirements Engineering Tools}.\hskip 1em plus 0.5em minus
  0.4em\relax Master Theses at Chalmers University of Technology, 2022,
  available online:
  \url{https://odr.chalmers.se/items/678b7cf4-9d5f-4ba0-a34e-4eda0965ecd8}.

\bibitem{TRA}
O.~Gotel, J.~Cleland{-}Huang, J.~H. Hayes, A.~Zisman, A.~Egyed,
  P.~Gr{\"{u}}nbacher, A.~Dekhtyar, G.~Antoniol, J.~I. Maletic, and
  P.~M{\"{a}}der, ``Traceability fundamentals,'' in \emph{Software and Systems
  Traceability}, J.~Cleland{-}Huang, O.~Gotel, and A.~Zisman, Eds.\hskip 1em
  plus 0.5em minus 0.4em\relax Springer, 2012, pp. 3--22.

\bibitem{lauesen_2002}
S.~Lauesen, \emph{Software Requirements-Styles and Techniques}.\hskip 1em plus
  0.5em minus 0.4em\relax Addison Wesley, 01 2002.

\end{thebibliography}

\newpage

\appendix
\subsection{Instantiating the Extended Architectural Framework}
\begin{figure}
    \centering
    \includegraphics[width=0.5\textwidth]{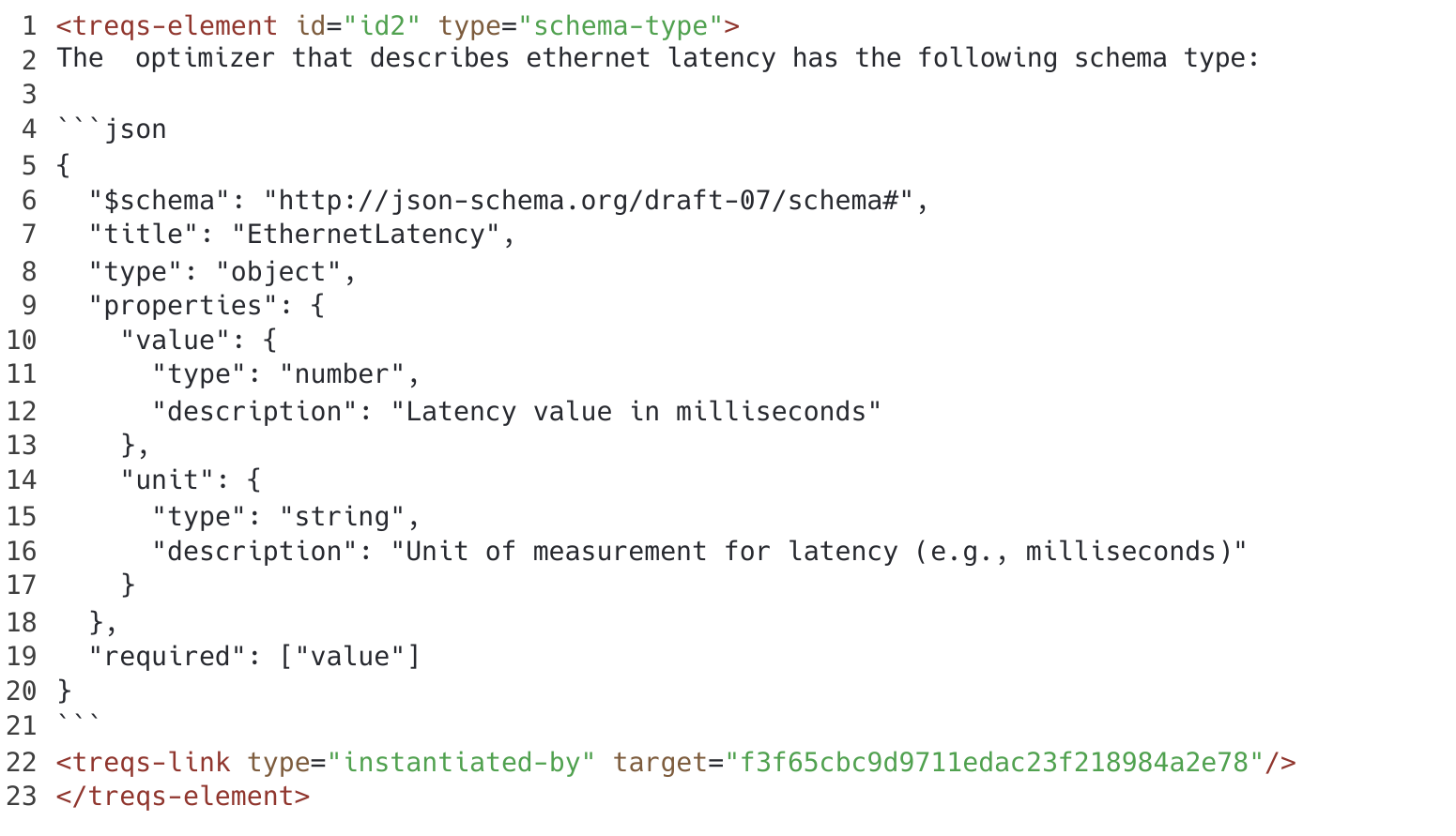}
    \caption{A \textit{schema-type} element that includes the JSON schema that describes an \textit{OptimizerInput} property related to \emph{ethernet latency}.}
    \label{fig:schema_type_ex}
\end{figure}

An important component of the extended architectural framework outlined in Figure \ref{fig:addition_to_caf}, is the \textit{schema-type}. An example of which is shown in Figure \ref{fig:schema_type_ex}.
\subsection{Target System Configuration Schema}
\begin{figure}
    \centering
    \includegraphics[width=0.5\textwidth]{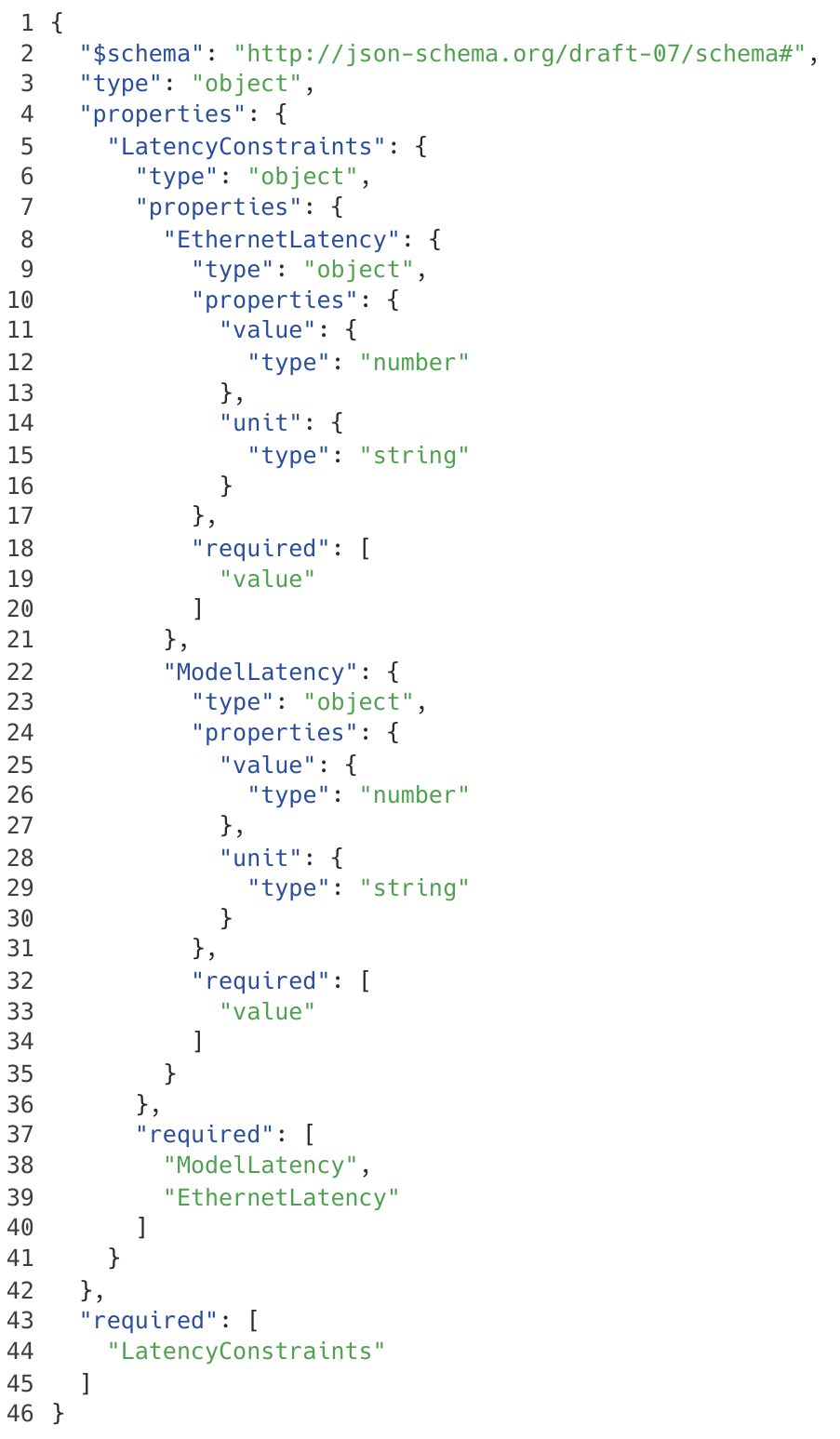}
    \caption{Input to GRAMS: A sample JSON schema describing the configuration of a target system that has two properties: \emph{ethernet latency} and \emph{model latency}.}
    \label{fig:sample_config_schema}
\end{figure}

The JSON schema for the configuration of the target system is given to GRAMS, not only for validation checking, but also to be included in the output of GRAMS. Figure \ref{fig:sample_config_schema} illustrates such a sample schema not only for the parts of the configuration that are to be populated from the requirement model, but for all properties. GRAMS traverses the schema and identifies and performs checks on the relevant properties.
\vspace{-1em}
\subsection{Intermediary Output Format}
The intermediarey output format is YAML and it has two main sections:
\begin{itemize}[leftmargin=3mm] 
    \item \textit{config\_schema} defines a JSON schema for configuring the target system, specifying schemas for extracting properties from the architectural framework. These properties may be nested, and the schema serves as input to GRAMS. Additionally, it's included in the output, enabling direct use of the intermediary output by the target system or by a tool for specific adjustments.
    \item \textit{optimizer\_inputs} contains a list of elements specifying \textit{OptimizerInput} properties, for each element its \textit{file\_name}, \textit{label}, \textit{placement}, \textit{treqs\_type}, and \textit{uid}, along with traceability path pointing to requirements going all the way up to a \textit{runtime\_scenario}. Additionally, it includes a schema defining the JSON schema for the property and value indicating the specific constraint in milliseconds. This structure repeats for each requirement.
\end{itemize}
\subsection{GRAMS Output}
\begin{figure*}
    \centering
    \includegraphics[width=\textwidth]{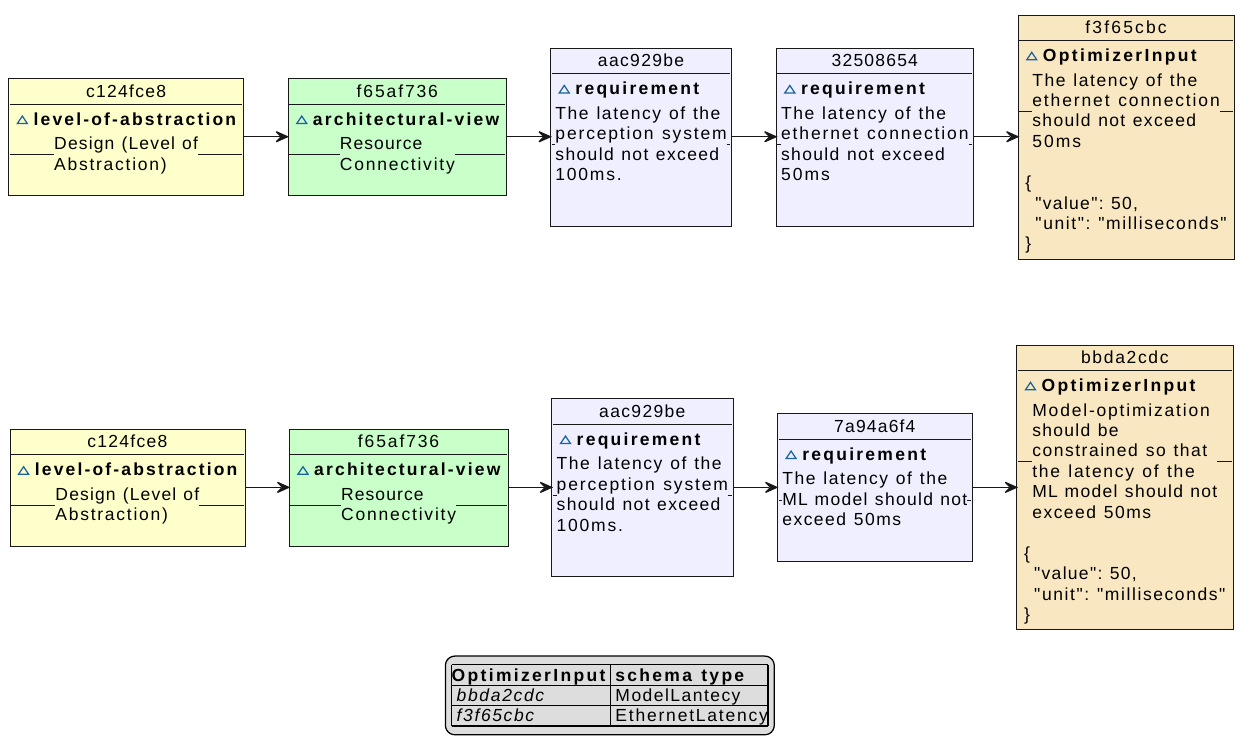}
    \caption{Rendering of GRAMS PlantUML output which illustrates two paths stemming from the same \textit{runtime-scenario} down to an \textit{OptimizerInput} property. The legend shows the types of the two identified properties.}
    \label{fig:sample_plantuml}
\vspace{-1em}
\end{figure*}

GRAMS output can be displayed in two formats: PlantUML or YAML, the latter is intended for consumption of the target system (or other intermediary tooling) whereas the PlantUML version is for providing an overview. A sample \href{https://gitlab.com/latiif/configgen/-/snippets/3690884}{PlantUML} and \href{https://gitlab.com/latiif/configgen/-/snippets/3690883}{YAML} outputs are hosted as snippets, whereas Figure \ref{fig:sample_plantuml} is the visualization on the PlantUML code. A \href{https://asciinema.org/a/648823}{recording} of GRAMS' YAML output is available.

\subsection{GRAMS Checks}
As outlined in \ref{sec:checks}, the following recordings demonstrate how GRAMS detects and reacts to the various checks:
\begin{itemize}
    \item \href{https://asciinema.org/a/648831}{Meta-model inconsistencies}
    \item \href{https://asciinema.org/a/648832}{Internal schema correctness}
    \item \href{https://asciinema.org/a/648834}{Semantic equivalence between configuration schema and framework's schema}.
\end{itemize}

\end{document}